# Full simulation of the open field lines

# above pulsar's polar cap – Part I

## Yudith Barzilay

Physics Department, Ben-Gurion University P.O.B. 653, Beer-Sheva 84105, Israel;  yudith@bgu.ac.il

**Abstract**

We have programmed a full simulation of the open field lines above the polar cap of a magnetized pulsar, using time dependent Particle In Cell (PIC) numerical code (Birdsall and Langdon 1991). We consider the case of free ejection of electrons from the star surface, a Space Charge Limited Flow (SCLF) model. We report here the first results of the simulation. Electrons are accelerating along the open field lines to the flat space-time SCLF maximum Lorentz factor prediction, with oscillation pattern. Than we add the General Relativistic (GR) Frame Dragging correction that provide the particles the high Lorentz factor ($10^6$-$10^7$) needed to initiate pair production. The electrons accelerate according to the analytic expressions given in Muslimov and Tsygan 1992 and Muslimov and Harding 1997, with   oscillation pattern. Electron-positron pair production is now being programmed, using the cross sections appears in the literature, and Monte-Carlo code. After completing this stage, we will **automatically** get: a) the positron return current (thus we could calculate the polar cap heating and the X-ray emission). b). The photons and electrons observed spectrum (photons and electrons that escape the magnetosphere after the cascade). c). The pulsar death line (pulsars with not enough pair production). d). The PFF height (pair formation front location). Those results will be report in a different paper.

## 1).  Introduction

We consider the basic model for pulsar activity, the case of an electron's beam extracted from the star surface above the polar cap region. The electrons accelerate along the open field lines due to the electric field induced by the rotation of the magnetized pulsar, as Space Charge Limited Flow (SCLF), as described by Michel 1974, Fawley, Arons and Scharlemann 1977, Cheng and Ruderman 1977, and Arons and Scharlemman 1979 (hereafter M74, FAS77, CR77, AS79 respectively). Frame dragging, a General Relativity (GR) effect, enables to accelerate the electrons to much higher energies as was found by Muslimov and Tsygan 1992 and Muslimov and Harding 1997 (hereafter MT92, MH97 respectively). Those primary relativistic electrons initiate cascade of pair production. The initial calculations of pair production consider only pair production via Curvature Radiation (CR) (AS79). Now Inverse Compton Scattering (ICS) and Synchrotron Radiation (SR) are also considered as important mechanisms, for both particles energy loss and pair production.

In this paper we report the particle acceleration results along the magnetic field lines using Particle In Cell (PIC) numerical code presented by Birdsall and Langdon 1991 (hereafter BL91), first without the GR Frame dragging correction, and than with the GR correction. The pair cascade results including CR, ICS, SR using the cross sections appears in the literature, and Monte-Carlo code will be report in a different paper, with the cascade consequence as: the positron return current (thus we could calculate the polar cap heating and the X-ray emission), the photons and electrons observed spectrum (photons and electrons that escape the magnetosphere after the cascade), the pulsar death line (pulsars with not enough pair production) and the PFF height (pair formation front location).

The model and the equations are introduced in section 2, numerical results are presented in section 3, and discussion appears in section 4.

## 2). The Model and the Equations

We consider a cylindrical conductive tube with constant radius $R$ (which is the polar cap radius), with background Goldreich-Julian charge density $\rho_{GJ}$. The tube has constant radius $R$, since the influence of the decline of the magnetic field due to field line diverge is equal for the charge density $\rho$ and $\rho_{GJ}$, and only the difference ($\rho$ - $\rho_{GJ}$) should be consider. Every time step, electrons are continuously inserted from the base of the tube (representing the star surface) as long as $\rho_{GJ}$ at the base of the tube is not compensated. The tube is represented in the simulation as discrete cells on a grid. The fields are discrete at each cells, while the particles positions and velocity are continues, and weighting (interpolation) between the discrete and continues values is done, i.e. PIC numerical code, as presented by BL91. The fields are calculated from Maxwell's equation by knowing the position and velocities of all the particles. The forces on the particles are found using the electric field, and the positions and velocities of the particles are found using the equations of motion. This is repeated at each time step. Figure 1 shows one time step in the simulation (not including pair production).

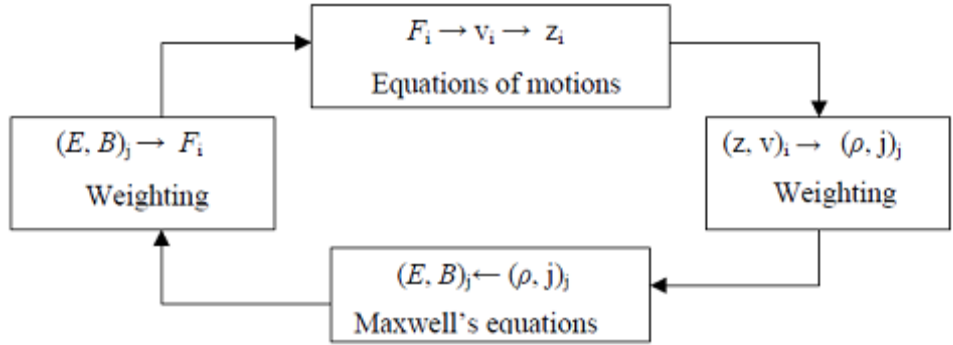

Figure A – one time step in the simulation. $F_i$ $v_i$ $z_i$ are the force, velocity and position of the $i$'th particle, $E_j$ $B_j$ $\rho_j$ $j_j$ are the electric and magnetic fields, the charge density and current at each grid cell $j$.

The equation of motion:

$$F = qE_z = \frac{dp}{dt} = \frac{d(\gamma mv)}{dt} \qquad , \qquad v = \frac{dz}{dt} \qquad (1)$$

where $\gamma$ is the Lorentz factor, $z$ is height above the surface, $q$ is the electron charge.

Maxwell's equations (FAS77):

$$\nabla \cdot E = 4\pi(\rho - \hat{\rho})$$
$$\nabla \cdot B = 0$$
$$\nabla \times B = \frac{4\pi}{c}(j - \hat{j}) + \frac{1}{c}\frac{\partial E}{\partial t} \qquad (2)$$
$$\nabla \times E = -\frac{1}{c}\frac{\partial B}{\partial t}$$

Well inside the light cylinder then $\hat{\rho} = \rho_{GJ}$ , $\hat{j} = 0$.

Assuming conductive walls and transverse derivatives $\partial/\partial r \rightarrow 1/R$ we rewrite those equations as:

$$\frac{1}{c}\frac{\partial E_z}{\partial t} = \frac{B_\varphi}{R} - \frac{4\pi}{c}j$$
$$\frac{1}{c}\frac{\partial E_r}{\partial t} = -\frac{\partial B_\varphi}{\partial z} \qquad (3)$$
$$\frac{1}{c}\frac{\partial B_\varphi}{\partial t} = -\frac{\partial E_r}{\partial z} - \frac{E_z}{R}$$

We write those equations in the leap-frog method, as in Lyubarsky 2009 (for Maxwell's equations).
We initiate the electric fields by solving Poisson equation (FAS77) with empty tube ($\rho$=0):

$$\nabla^2 \phi = \nabla^2_{\parallel} + \nabla^2_{\perp} = \frac{d^2\phi}{dz^2} - \frac{\phi}{R^2} = -4\ \pi(\rho - \rho_{GJ}) \tag{4}$$

where $\Phi$ is the electric potential. Since the tube is initially empty, the magnetic field $B_\varphi(t=0)=0$.

The boundary conditions for Poisson equation are: $\Phi = 0$ on the bottom of the tube (the star surface), $\Phi' = 0$ at the end of the tube. The additional boundary condition in the tube $\Phi = 0$ on the tube walls is satisfied automatically from the way we rewrite Maxwell's equations due to the assumption of conductive walls.

**We get** the SCLF condition $\Phi' = 0$ at the bottom of the grid (the star surface) as the simulation run. **We also get** a cloud of electrons at the bottom of the grid as expected in SCLF model.

The background Goldreich-Julian charge density $\rho_{GJ}$ should be initiated too. For flat space-time $\rho_{GJ}$ is presented in AS79:

$$\rho_{GJ} = -\frac{\Omega_* B_*}{2\pi\ c}\left(\frac{R_*}{r}\right)^3\left[\cos i + \frac{3}{2}\sqrt{\frac{r}{R_*}}\theta_* \sin i\ \cos\varphi\right] \tag{5}$$

where $i$ is the inclination angle (angle between magnetic and rotation axes), $r = R_*+z$, , $R_*$ is the star radius, $\varphi$ and $\theta_*$ are the magnetic colatitude and azimuth angle of a field line where it intercepts the stellar surface ( $\varphi = 0$, $\theta_* = \xi\ \theta_p = 0.67\ \theta_p$ chosen to be the field line were the plasma flow is centered (AS79) ). $\theta_p$ is the polar cap angle, $\Omega_*$ is the star angular velocity, $B_*$ is the magnetic field at the star surface, and c is the speed of light. The second term of equation 4 is due to the opening of the magnetic field lines with height, and is vanishing for an aligned rotator. We dropped the factor $(R_*/r)^3$, since in our model the tube has constant radius $R$.

In the simulation, the above $\rho_{GJ}$ was inserted into the grid, and than Poisson equation is solved to initiate the fields. The simulation run as was described before: each time step all the particles location and velocities are being update, as well as the fields at the entire grid. The program result is compared to the analytic expression given in M74.

MT92, MH97 have shown that General Relativity (GR) frame dragging is very important for particles acceleration. They derive $\rho_{GJ}$ in curved space-time (which reduce to AS79 expression in flat space-time):

$$\rho_{GJ} = -\frac{\Omega_* B_*}{2\pi\ c\alpha}\left(\frac{R_*}{r}\right)^3\frac{f(\eta)}{f(1)}\left[\left(1 - \frac{\omega}{\Omega_*}\right)\ \cos i + \frac{3}{2}\left(\frac{r}{R_*}\right)^{\frac{1}{2}}\frac{H(\eta)}{f(\eta)^{1/2}}\theta_* \sin i\ \cos\varphi\right] \tag{6}$$

The GR information is stored in the two variables $\alpha$ and $\omega$. $\alpha = (1 - r_g/r)^{1/2}$ is the gravitational redshift, $r_g = 2GM_*/c^2$ is the gravitational radius ($r_g/R_* \sim 0.4$), $\omega$ is the modification to the star's angular velocity due to frame dragging effect. Here we modify $\rho_{GJ}$ by replacing $\Omega_*$ with $\Omega_* - \omega$.

$$\omega = \Omega_*\ \frac{2}{5}\left(\frac{r_g}{R_*}\right)\left(\frac{R_*}{r}\right)^3 = \Omega_*\ \frac{\kappa}{\eta^3} \tag{7}$$

where $\eta = r/R_*$ the functions $f(\eta)$ and $H(\eta)$ are given in MT92, MH97, and are numerical factors for curved space-time, which are equal to 1 in flat space-time. The $\sin i$ term is comparable in flat and curved space-time, and much smaller than the $\cos i$ term near the star. Since the $\cos i$ term is much significant near the star, pulsars with small inclination angle, accelerate particles to higher energies.

Full GR treatment includes modifying Maxwell's equation and the equation of motion with the factor $\alpha$ as well, as describe in MT92, MH97, and additional effects describe in Gonthier and Harding 1994, estimated there to result in ~10% corrections. As before, we drop the field line curvature effect in $\rho_{GJ}$.

In the simulation, the expression of equation 6 was inserted into the grid instead the expression of equation 5, and than Poisson equation is solved to initiate the fields, and the simulation run as was described before. The program result is compared to the analytic expression:

$$\gamma_{max} = \frac{e\ \phi}{mc^2} + 1 \tag{8}$$

where $\Phi$ is given in MT92, MH97 (the field lines curvature effect in $\Phi$ should be removed).

## 3). Results

The result is this paper are for a "standard" pulsar of 1sec period, magnetic field of $B=10^{12}$ G, star radius $R_*=10^6$ cm, polar angle $\theta_p=\sqrt{(\Omega_* R_*/c)}=0.014$, polar cap radius $R_p = R_* \theta_p \sim 10^4$ cm, star mass $1.4 M_{sun}$. The Goldreich - Julian charge number density at the star surface is $n_{GJ}=6.6*10^{10}$ *$\cos i$ cm$^{-3}$ ($\rho_{GJ} = n_{GJ} q = \Omega_* B \cos i / 2\pi c$). The inclination angle $i$ was taken to be $i=0$ for the runs of flat space-time SCLF, and $i=0.5$ rad$\sim 28^o$ for the GR frame dragging correction. All those values are parameters at the simulation and can be change for each run of the simulation.

The simulation values $n_{GJ}$ (charge number density), $q$ (electron charge), $m$ (electron mass) are normalized according to the plasma frequency $\omega_p = q \sqrt{4\pi\ n_{GJ} / m}$. Distances are normalized according to the polar cap/tube radius $R$. The speed of light in the simulation is taken to be c=1.

The first part of the results, is a run of flat space-time SCLF. This step provide a relation between "real world" values and simulation values. It is only to "calibrate" the simulation values with the pulsar values, i.e. to indicate the values of $(R\ ,\ \omega_p)$ that should be taken in the simulation, to represent the pulsar values.

Table 1 and Figure 2 show results of different tube radius $R$ and different plasma frequency $\omega_p$, such that the maximum Lorentz factor $\gamma_{max}$ remains constant. The two last red columns of Table 1 are for "real world". The simulation result (blue line) is compared with analytic expression given in M74 (red line). The simulation result shows a "front". Behind the front the particles reaches the maximum Lorentz factor expected by SCLF, with oscillatory pattern. The results of Figure 2 are all taken when the particles reaches 30$R$. Moving from the left panel to the right panel ('a'   'b'   'c'), $R$ is increases (3,000  5,000  10,000 respectively) thus it takes more time step 't' (90,000  150,000  300,000 respectively), and more particles involved (78,161  130,237  260,263 respectively), while the pattern look the same. Here negative Lorentz factor indicate returning electrons, i.e. electrons that move backwards.

|  | Program $R$=3,000 | Program $R$=5,000 | Program $R$=10,000 | Program $R$=3,000 | Real World $R_p \sim 10^4$ cm |
|---|---|---|---|---|---|
| $\gamma_{max} = \sqrt{2}\ \omega_p\ R / c$ | ~15 | ~15 | ~15 | ~7,000 | ~7,000 |
| $\omega_p = q\sqrt{4\pi\ n_{GJ} / m}$ | 0.0035 | 0.0021 | 0.0011 | 1.66 | $1.5*10^{10}$ s$^{-1}$ |

<u>Table 1</u>  Analytic results for flat space-time SCLF with same $\gamma_{max}$.

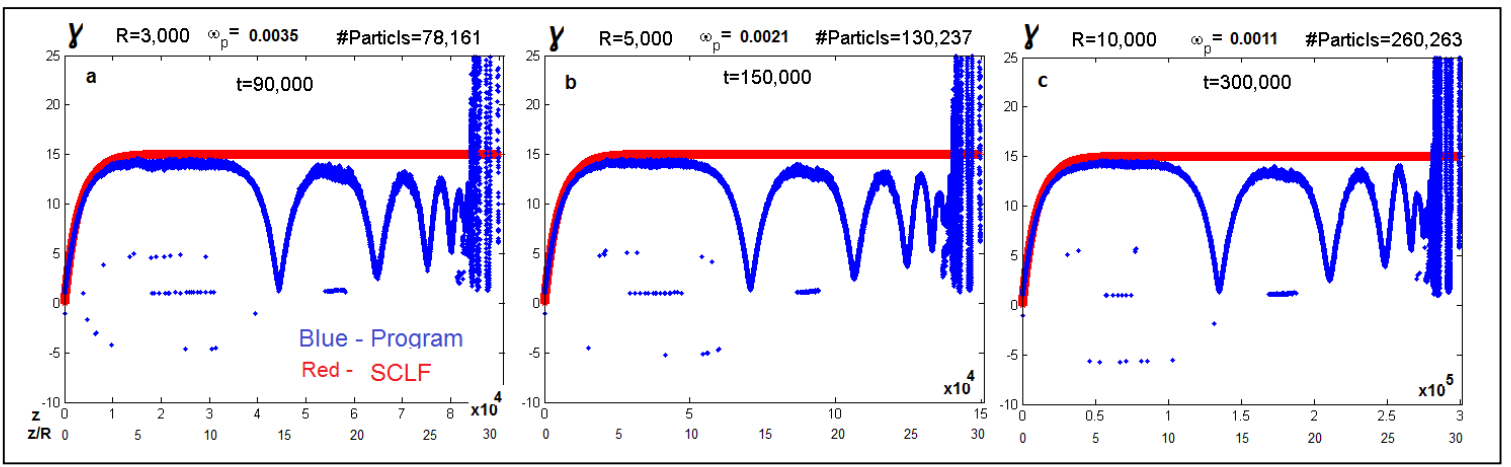

<u>Figure 2</u>  Program results for flat space-time SCLF with same $\gamma_{max}$. The blue line is the program results, the red line is the analytic flat space-time SCLF presented in M74.

Table 2 and Figure 3 show results of different tube radius $R$, while $\omega_p$ is constant. The two last red columns in Table 2 are for "real world". Again, the results are taken when the particles reaches 30$R$. Here again $R$ is increases from the left panel to the right panel, results in more time-step and more particles involved. As $R$ is increasing while $\omega_p$ is constant (approaching "real world" values), $\gamma_{max}$ is increases too, and the shape becomes more oscillatory.

| | Program $R$=3,000 | Program $R$=5,000 | Program $R$=10,000 | Program $R$~5.0e6 | Real World $R_p \sim 10^4$ cm |
|---|---|---|---|---|---|
| $\gamma_{max} = \sqrt{2} \; w_p R / c$ | ~5 | ~8 | ~15 | ~7,000 | ~7,000 |
| $\omega_p = q\sqrt{4\pi \; n_{GJ} / m}$ | 0.0011 | 0.0011 | 0.0011 | 0.0011 | $1.5*10^{10}$ s$^{-1}$ |

Table 2   Analytic results for flat space-time SCLF with varying $R$, equal $\omega_p$.

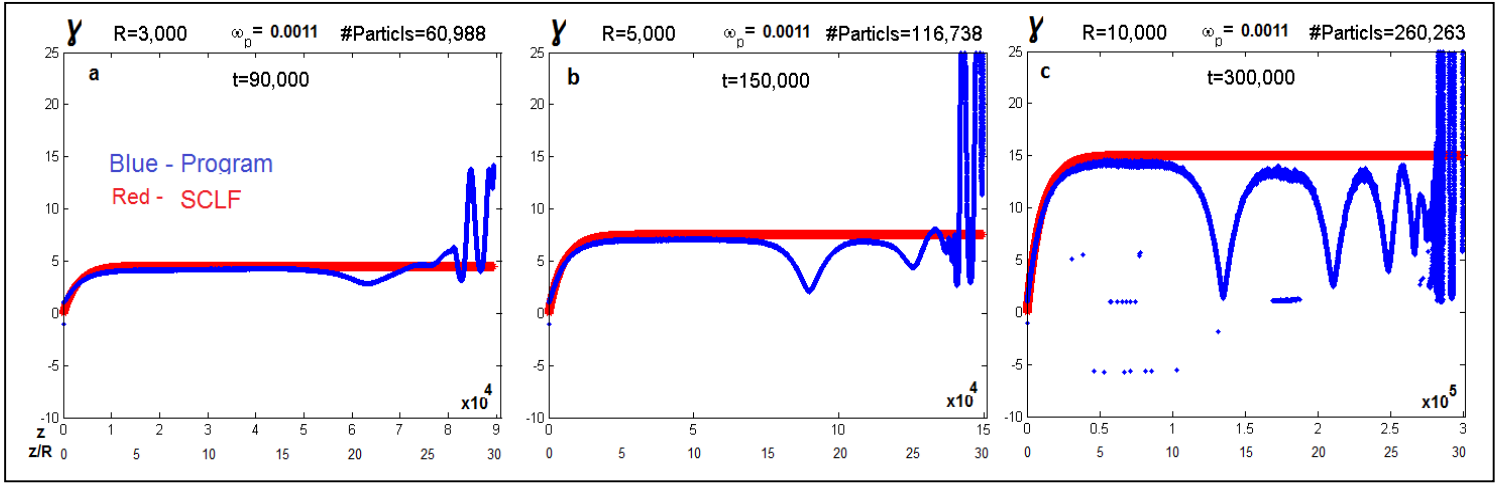

Figure 3   Program results for flat space-time SCLF with varying $R$, equal $\omega_p$.

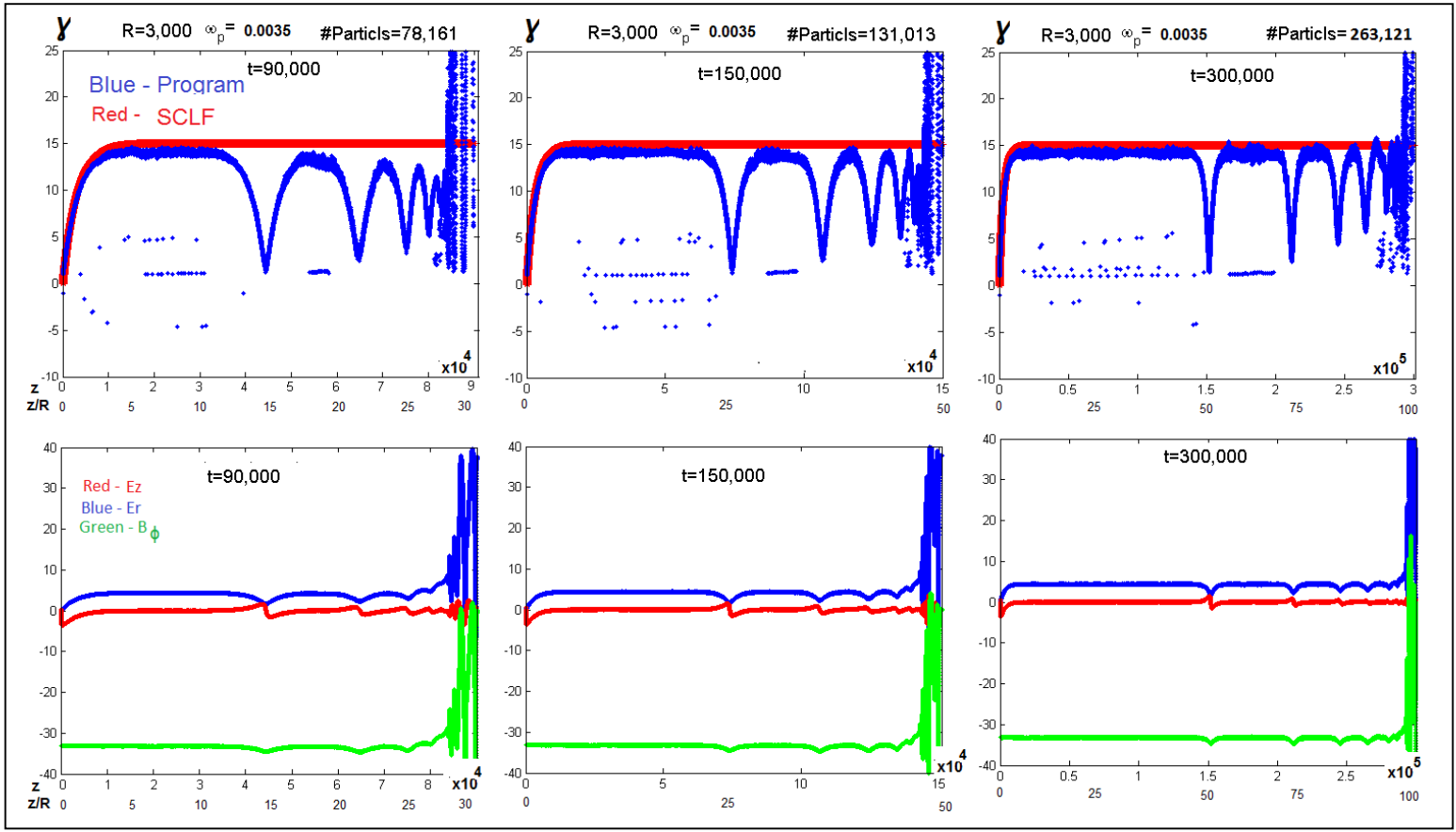

Figure 4   Program results for flat space-time SCLF with equal $R$, equal $\omega_p$, at different distance from the bottom of the tube (the star surface). The corresponding fields appear below.

Figure 4 shows results of the same ($R$ , $\omega_p$), with different distance from the bottom of the tube (the star surface). The results are taken for 30$R$, 50$R$, 100$R$. Also it takes more time-step, and more particles are involved, the pattern looks the same. The bottom panels show the corresponding fields. The red line is the parallel accelerating electric field $E_z$, the blue line is $E_r$, and the green line $B_\varphi$.

Note the SCLF condition that appears at the bottom of the tube, i.e. the vanishing of the parallel electric field at the star surface.

The second step in the simulation is runs with the GR frame dragging correction.

Figure 5 show results of the "standard pulsar" parameters describe above. Two different possibility of the pair ($R$ , $\omega_p$) that fit the pulsar parameters are shown (6,253  1.1) and (10,317  0.66) respectively, using the SCLF calibration. The results are taken at two different distances from the bottom of the tube: ~43$R$ (upper panels) and ~53$R$ (lower panels). Since $R$ is increases from the left panels to the right panels, it take more time-step and more particles are involved in the right panels, but the figures look the same (the two upper panels, and the two lower panels). As can be seen, there is a "front". Behind the front the particles have the Lorentz factor expected by the GR frame dragging correction. The figures show oscillatory pattern, with increasing oscillations as the distance from the star is increasing. The blue line is the program results, the green line is the analytic GR frame dragging correction presented in MT92, MH97, the red line is the analytic flat space-time SCLF presented in M74.

Figure 6 shows results of the same run as Figure 5, but with smaller inclination angle $i$=0.1 rad, resulting with acceleration to higher energies. Thick green line is the analytic expression for $i$=0.1 rad, thin green line is the analytic expression for $i$=0.5 rad.

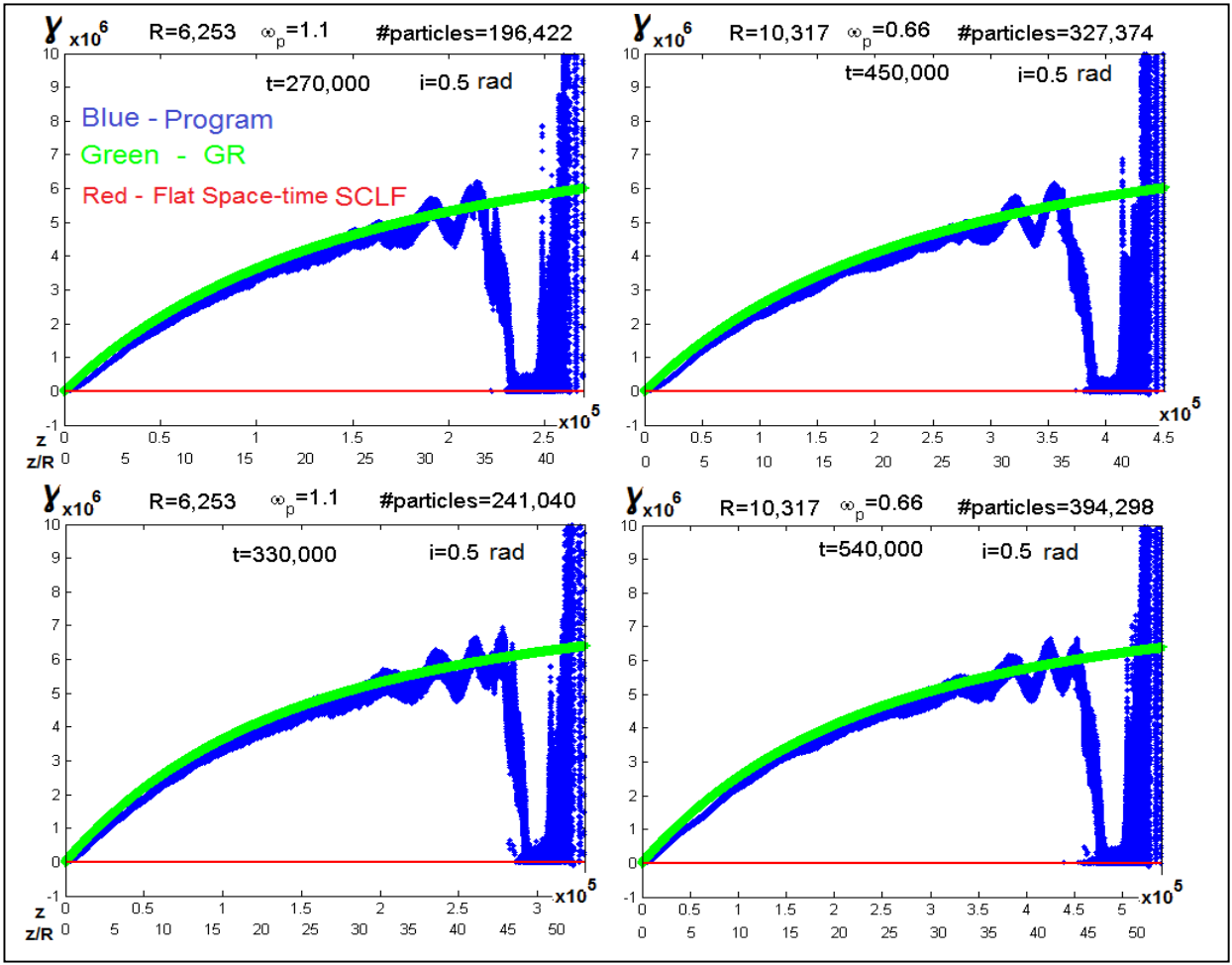

Figure 5  Lorentz factor with GR frame dragging effect. The upper panels are for distance ~43$R$ from the bottom of the tube (the star surface), the lower panels are for distance ~53$R$ from the bottom of the tube. One simulation time step for the left panels is equivalent to $3.8*10^{-11}$ seconds, and for the right panels to $2.3*10^{-11}$ seconds.

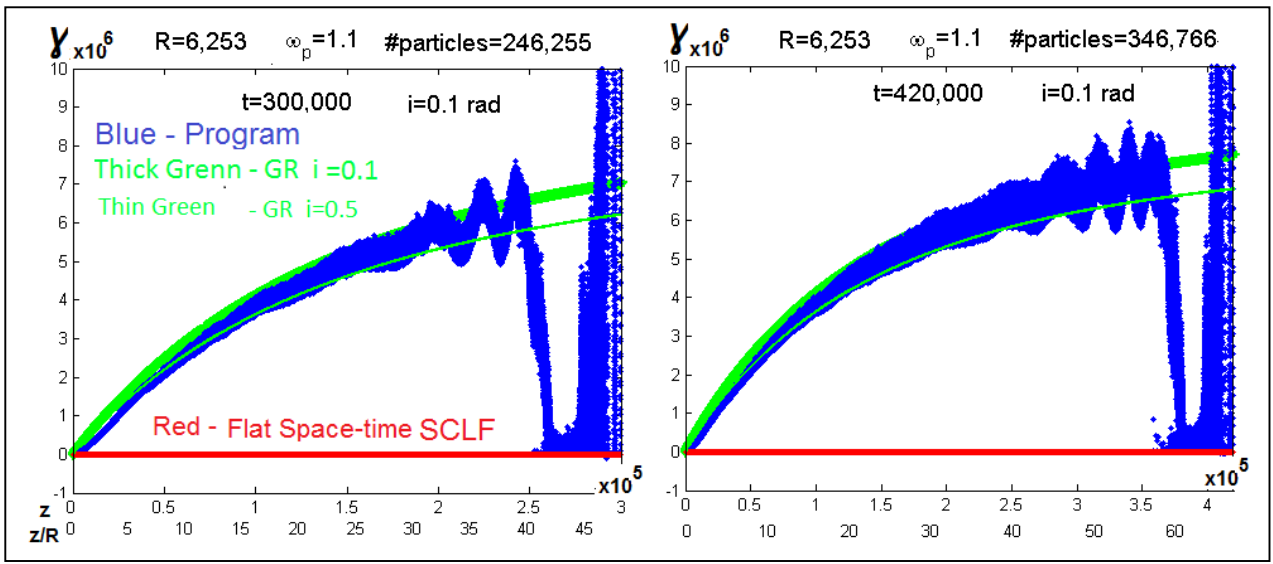

Figure 6 Same as Figure 5, but with smaller inclination angle. Two different times are shown.

## 4).  Discussion and Conclusions

We have programmed full simulation of particle acceleration above the open field lines of a pulsar's polar cap, using equation of motion and Maxwell's equations in PIC numerical code.

The first part is a run with flat space-time SCLF as presented in M74, FAS77, CR77, AS79. We got the SCLF condition at the bottom of the tube, i.e. the vanishing of the parallel electric field at the star surface. We also got a cloud of electrons at the bottom of the grid (the star surface) as expected for SCLF model. This step enables to calibrate the program distances and $n_{GJ}$ , $q$ , $m$ to the "real world" pulsar's values, through the tube/polar cap radius $R$ and the plasma frequency $\omega_p$. This step shows increasing oscillatory pattern as the program parameter $R\,\omega_p$ /c increases, resulting eventually with appearance of instability.

The second step is run with the GR frame dragging correction presented in MT92, MH97. The electrons are accelerated according to the analytic estimates, reaching Lorentz factor of order $10^6$ . The particles Lorentz factor shows oscillatory pattern that increases with distance from the star.

The third step will be reported in a next paper, and will include pair production. Primary electrons accelerating along the open field lines emit CR photons, or upscatter thermal photons from the star via ICS. The CR or ICS photon can produce a pair via magnetic pair production $\gamma B \rightarrow e^- e^+ B$. If one of the pair members is at high Landau level, than it radiate its excess energy via SR. The SR photon can also produce a pair. Thus cascade of CR / ICS / SR photons and pairs is produced. After completing this stage, we could report as well the positron return current (thus we could calculate the polar cap heating and the X-ray emission), the photons and electrons observed spectrum (photons and electrons that escape the magnetosphere after the cascade), the pulsar death line (pulsars with not enough pair production) and the PFF height (pair formation front location).

I thank Eduardo Guendelman for useful discussions, valuable suggestions and encouragement.